\documentclass[aps,prl,twocolumn,showpacs,superscriptaddress,reprint]{revtex4-1}
\usepackage{graphicx}
\usepackage{natbib}
\usepackage{amsmath}
\usepackage{amssymb}
\usepackage{bm}
\usepackage{color}
\usepackage{float}

\begin{document}

\title{Extremely low-energy collective modes in a quasi-one-dimensional system}
\author{Z. X. Wei} 
\affiliation{State Key Laboratory of Electronic Thin Films and Integrated Devices, University of Electronic Science and Technology of China, Chengdu 610054, China}
\author{S. Zhuang}
\affiliation{Beijing National Laboratory for Condensed Matter Physics, Institute of Physics, Chinese Academy of Sciences, Beijing 100190, China}
\author{Y. L. Su} 
\affiliation{State Key Laboratory of Electronic Thin Films and Integrated Devices, University of Electronic Science and Technology of China, Chengdu 610054, China}
\author{L. Cheng} 
\affiliation{State Key Laboratory of Electronic Thin Films and Integrated Devices, University of Electronic Science and Technology of China, Chengdu 610054, China}
\author{H. D. Zhou}
\affiliation{Department of Physics and Astronomy, University of Tennessee, Knoxville, Tennessee 37996, USA}
\author{Z. Jiang}
\affiliation{School of Physics, Georgia Institute of Technology, Atlanta, Georgia 30332, USA}
\author{H. Weng}
\affiliation{Beijing National Laboratory for Condensed Matter Physics, Institute of Physics, Chinese Academy of Sciences, Beijing 100190, China}
\author{J. Qi}
\email{jbqi@uestc.edu.cn} 
\affiliation{State Key Laboratory of Electronic Thin Films and Integrated Devices, University of Electronic Science and Technology of China, Chengdu 610054, China}

\date{\today}

\begin{abstract}
We have investigated the quasiparticle dynamics and collective excitations in the quasi-one-dimensional material ZrTe$_5$ using ultrafast optical pump-probe spectroscopy. Our time-domain results reveal two coherent oscillations having extremely low energies of $\hbar\omega_1\sim$0.33~meV (0.08 THz) and $\hbar\omega_2\sim$1.9~meV (0.45 THz), which are softened as the temperature approaches two different critical temperatures ($\sim$54~K and $\sim$135~K). We attribute these two collective excitations to the amplitude mode of charge density wave instabilities in ZrTe$_5$ with tremendously small nesting wave vectors. Furthermore, scattering with the $\hbar\omega_2$ mode may result in a peculiar quasiparticle decay process with a timescale of $\sim$1-2 ps below the transition temperature $T^*$ ($\sim$135~K). Our findings provide pivotal information for studying the fluctuating order parameters and their associated quasiparticle dynamics in various low-dimensional topological systems and other materials. 
\end{abstract}
\maketitle

In a many-body system, interaction between the quasiparticles, and/or interaction between the quasparticles and other quantized modes may bring about various broken symmetry ground states \cite{G book 1994}, e.g. superconducting state, density wave state, and magnetically ordered state, accompanied simultaneously by abundant collective excitations, typical examples of which are phonons, density waves (charge or spin) and magnons obeying the Bose-Einstein statistics. Delicate balance among different interactions and phases might cause the fluctuating characteristics in some order parameters or collective excitations with a finite correlation time $\tau_F$ \cite{G book 1994,Kivelson_RMP_2003}.  

ZrTe$_5$ is a quasi-one-dimensional material with highly anisotropic crystal lattice and electronic structures \cite{Weng_PRX_2014}. Their subtle correlations \cite{Weng_PRX_2014,J sciadv 2019,C PRX 2020} make this material enormously interesting. On the one hand, ZrTe$_5$ is associated with the topological phases of matter, i.e. the topological insulator (TI) \cite{G PRL 2016,R PRX 2016,Weng_PRX_2014,Y PRB 2018,H PRB 2017,R PRX 2016} or the Dirac semimetal \cite{RY PRB 2015,Y NC 2016,Wang 2018}, which have been experimentally investigated by different equilibrium or quasi-equilibrium techniques, e.g. electronic transport \cite{Niu PRB2017,Terry PRB 1999}, infrared optical spectroscopy \cite{RY PRB 2015,ZG PNAS 2017,Jiang PRL2020,Jiang PRB 2017,B PRL 2018}, scanning tunneling microscope \cite{Li PRL 2016,R PRX 2016}, and angle-resolved photoemission spectroscopy (ARPES) \cite{R PRX 2016,H PRB 2017}. On the other hand, many unusual electronic responses such as the chiral magnetic effect \cite{Li_NPhy_2016}, anomalous Hall effect \cite{Liang NP 2018}, and 3D quantum Hall effect \cite{Tang Nature 2019} have been reported. 

It is believed that those exotic phenomena are connected to the peculiar fermionic and bosonic excitations in ZrTe$_5$ with the anomalous phase transition around temperature $T^*$, where the resistivity presents a peak \cite{FJ PRB 1981}. In particular, the static charge density wave (CDW) stabilized by the external magnetic field was employed to interpret the 3D quantum Hall effect at temperatures below 20 K \cite{Tang Nature 2019,Qin_PRL_2020}, while the collective Dirac polarons were proposed to explain the sharp resistivity peak around $T^*$ \cite{Fu_PRL_2020}. However, the fluctuating nature of any unique collective excitation(s) in ZrTe$_5$ has never been observed or elucidated, although the related dynamics could be critical to understand the emergent phenomena in this material previously observed via the quasi-equilibrium probing techniques. 

\begin{figure}
	\includegraphics[width=9cm]{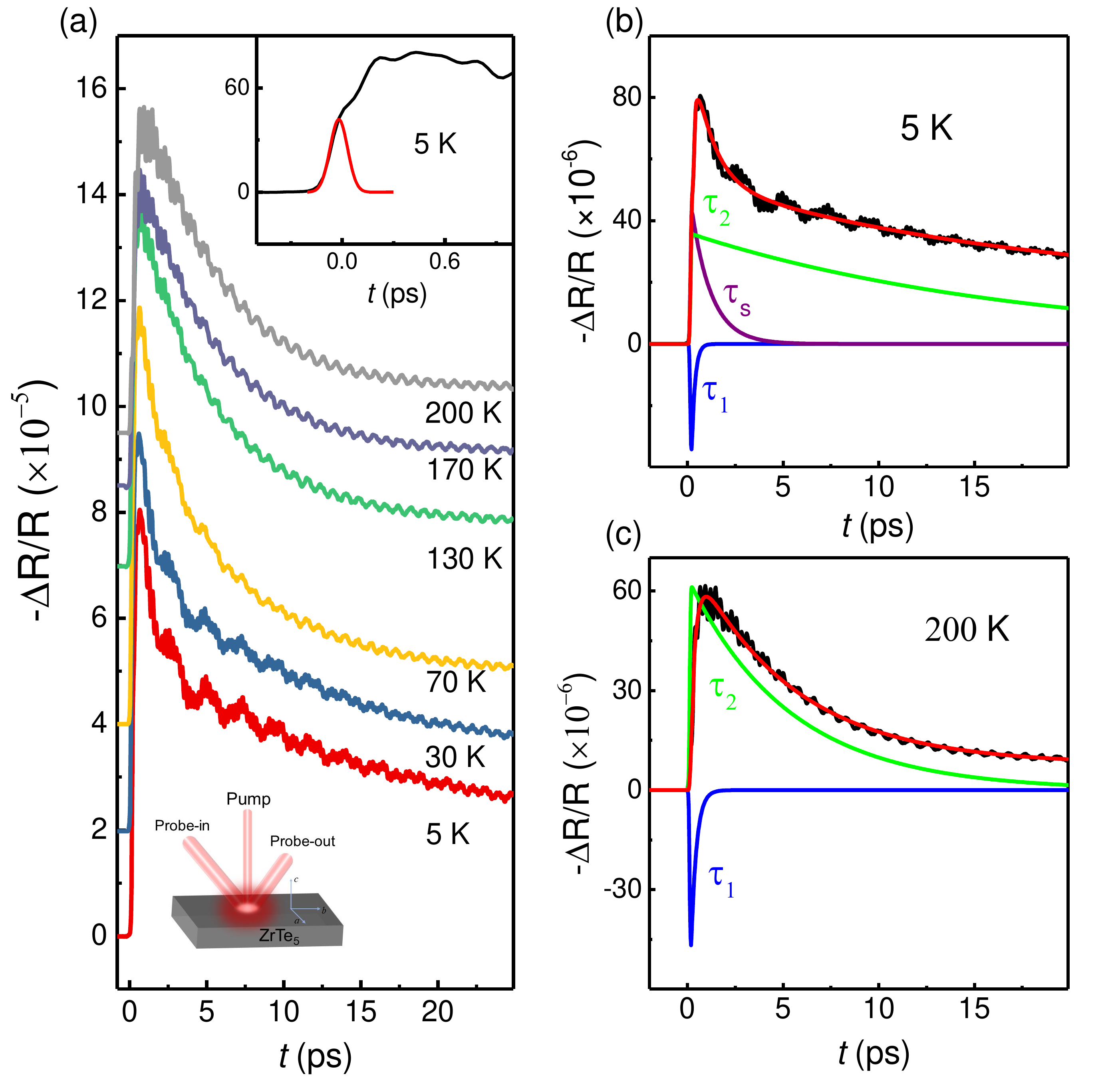}
	\caption{\label{fig:deltaR} (a) Typical $\Delta R(t)/R$ data as a function of temperature at a pump fluence of $\sim $0.5~$\mu$J/cm$ ^{2}$. Inset: $\Delta R(t)/R$ in short timescale. The red line is the pump-probe cross correlation profile. (b) and (c) Typical fitted curves (red solid lines) for $\Delta R(t)/R$ data using Eq.~(\ref{eq:deltaR_fit}) at 5 K and 200 K, respectively. The blue, green and purple lines are for components with decay times $\tau_1$, $\tau_2$, and $\tau_s$, respectively.}
	\vspace*{-0.2cm}
\end{figure}

By contrast, the ultrafast optical pump-probe technique is a powerful tool that not only can reveal the quasiparticle dynamics associated with the topological properties \cite{Gedik PRL2011,Wang PRL 2016} but also is capable of unveiling the collective phenomena in a broad correlation time- or length scale \cite{Yusupov_PRL_2008,Torchinsky_NM_2013,Liu PRL 2020}. Here, using ultrafast optical pump-probe spectroscopy our measurements for the first time reveal two extremely low-energy collective amplitude mode oscillations ($\sim$0.33 meV and 1.9 meV) at low temperatures in the high-quality ZrTe$_5$ single crystals \cite{note1}, which have a transition temperature $T^*$ of $\sim$135 K. We demonstrate these two modes arising from the fluctuating CDW triggered by the acoustic phonons. We also observe a novel quasiparticle dynamics associated with this type of bosonic mode emerging below $T^*$ with a ps-timescale. 

All measurements were performed using an ultrafast laser with a repetition rate of 80 MHz, a pulse width of $\sim$55 fs (FWHM), and a center wavelength tunable between 780 nm and 820 nm. Detailed information on experimental setup, sample preparation and additional data can also be found in the Supplemental Material \cite{Suppl} and Refs.~\cite{Wang PRL 2016, Qi PRL 2013,Liu PRL 2020,Jiang PRB 2017,Jiang PRL2020,W J 2016}. Figure \ref{fig:deltaR} (a) shows the typical measured signals at low and high temperatures. Upon photoexcitation, all the $\Delta R(t)/R$ signals clearly exhibit an instantaneous rise, succeeded by lateral relaxation processes. Besides, rich damped ultrafast oscillations are superimposed on the non-oscillating decay background. These coherent oscillations can verify the high quality of our samples, as will be discussed in detail later.      

\begin{figure}
	\includegraphics[width=8.5cm]{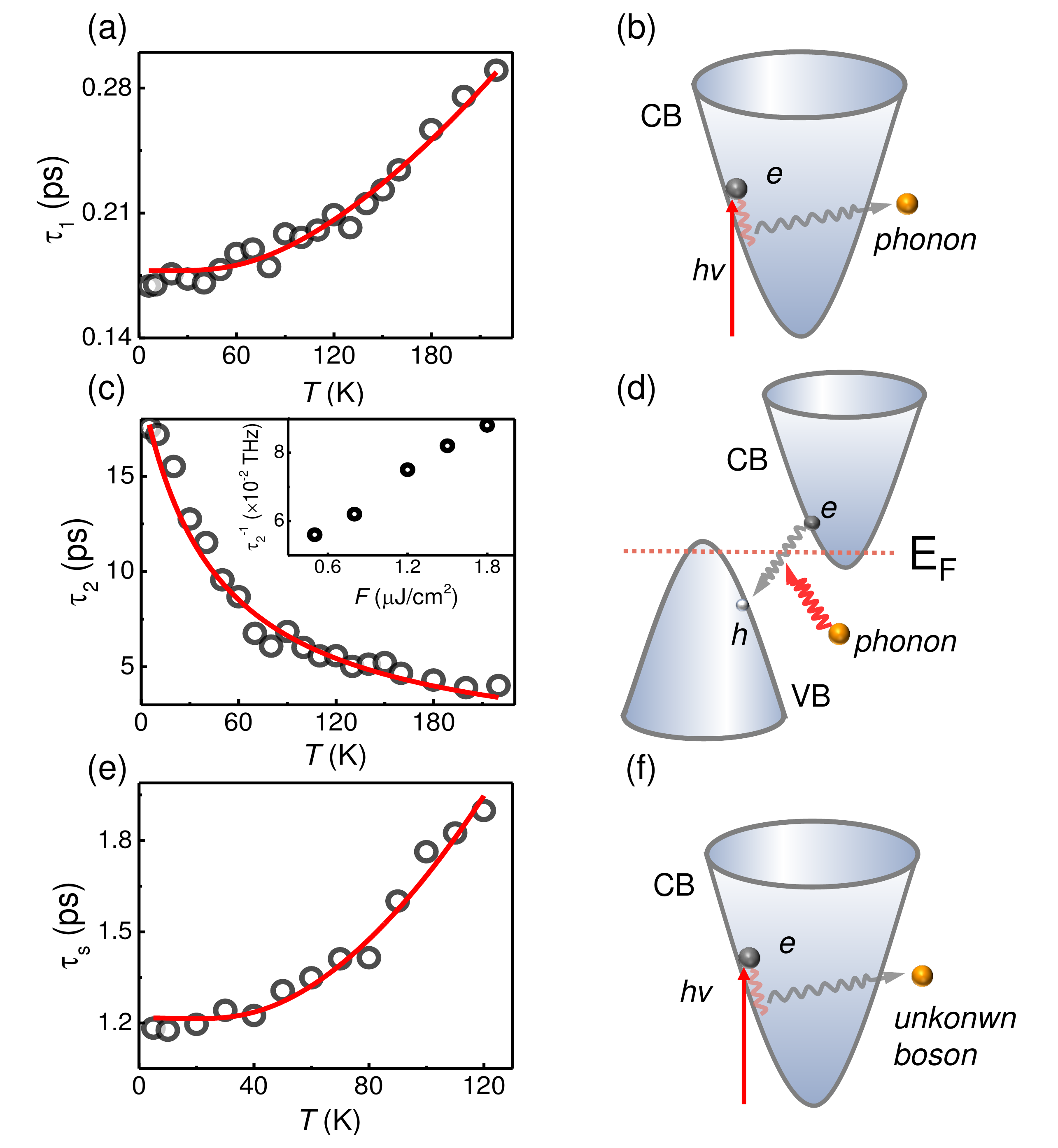}
	\caption{\label{fig:decay}  (a) $\tau_{1}$ as a function of temperature. Red line is a fit via TTM. (b) Schematic of the hot carrier excited by $h\nu$ photon cooling via e-ph scattering. (c) $\tau_{2}$ as a function of temperature. Red line is a fit based on the phonon-assisted e-h recombination process. Inset: $ 1/\tau_{2}$ as a function of pump fluence at 5 K. (d) Schematic of the phonon-assisted e-h recombination. (e) $\tau_{s}$ as a function of temperature. Red line is a fit via TTM.  (f) Schematic of the hot carrier cooling excited by $h\nu$ photon involved with unknown bosonic excitations emerging below $T^{*}$.}
	\vspace*{-0.2cm} 
\end{figure}

We first focus on the non-oscillatory signals. A close inspection of the rising curve (see the inset to Fig.~1(a)) reveals that it actually comprises two parts in the time-domain: (1) one exactly overlaps with the pump-probe cross correlation profile; (2) the other one shows relatively slow rising before the $\Delta R/R$ signal reaches the maximum value. We identify that the relative slow-rising can be well described by a sub-ps exponential decay process having amplitude with a different sign from that of the subsequent relaxation dynamics \cite{O PRL 2013, Qi APL 2010} (See also Supplemental Material for further details \cite{Suppl}). However, the succeeding relaxation process does not exactly follow a single exponential decay at all investigated temperatures. Rather, an extra decay process can be unambiguously observed below a critical temperature, which is nearly the same as the transition temperature $T^*\sim$135 K. According to these observations, we can fit our data using the following formula \cite{D PRL 2002,Wang PRL 2016},
\begin{equation}
	\Delta R(t)/R=(A_{1}e^{-\frac{t}{\tau_{1}}}+A_{2}e^{-\frac{t}{\tau_{2}}}+A_{s}e^{-\frac{t}{\tau_{s}}}+C)\otimes G(t),
	\label{eq:deltaR_fit}
\end{equation}
where $A_{j}$ and $\tau_{j}$ ($j=1,2,s$) are the amplitudes and decay times, respectively. $A_{1}>0$ and $A_{2},A_{s}\le0$. Here, $\tau_{1}$ and $\tau_{s}$ represent the sub-ps and extra decay processes, respectively. $\tau_{2}$ is the subsequent decay after $\tau_{1}$ or $\tau_{s}$. $C$ is a constant, and $G(t)$ is an Gaussian function standing for the pump-probe cross correlation. As seen in Figs.~\ref{fig:deltaR}(b) and (c), the fitted curves are in excellent agreement with the experimental data at low and high temperatures. Specifically, we found that $\tau_{1} $ is less than $\sim$0.3 ps, $\tau_{2}$ is $\sim$5-18 ps, and $\tau_{s}$ is $\sim$1-2 ps. The decay process characterized by $\tau_{s}$ emerges below $\sim 135$~K ($T^*$), or $A_s=0$ for $T\ge T^*$. The extracted $\tau_{j}$ ($j=1,2,s$) as a function of temperature are shown in Fig.~\ref{fig:decay}.

In general, the relaxation processes in conventional nonmagnetic materials disclosed by $\Delta R(t)/R$ can include the e-e thermalization, the e-ph scattering, the e-h recombination, and the thermal diffusion processes \cite{D Rev Mod Phys,J Book 2013,R 2002}. The e-e thermalization usually has a timescale less than $\sim$100 femtoseconds. If this process is faster than the excitation pulse and out of the time resolution limit, it is often concealed in the initial rising signal, which then follows the pump-probe cross correlation profile. This is well consistent with our experiments. On the other hand, the thermal diffusion process and the e-h recombination across a direct gap with radiative decay have timescales larger than $\sim$100 ps, and clearly do not contribute to the three processes characterized by $\tau_{j}$ ($j=1,2,s$) shown in Figs.~\ref{fig:deltaR} and \ref{fig:decay}. 

Since $\tau_{1}$ and $\tau_{2}$ decay processes exist at all investigated temperatures, and experience a continuous change across $T^*$, they should represent a generic dynamical responses of ZrTe$_5$, independent of any physics underlying the transition around $T^*$. In fact, the sub-ps process characterized by $\tau_{1}$ can be attributed to the photoexcited hot-carrier cooling via e-ph scattering (see Fig.~\ref{fig:decay}(b)), which in turn leads to the optical phonon emission. Its $T$-dependence is quite similar with those of e-ph thermalization times observed in metals, semiconductors and topological materials  \cite{Liang APL 2014,M PRB 2005,R PRB 1995,Qi APL 2010,J Book 2013,M Nano Lett 2012}. Such $T$-dependent behavior can be described by the two-temperature model (TTM) \cite{Allen PRL 1987,R PRB 1995}. Fig.~\ref{fig:decay}(a) demonstrates that the fitted results using TTM agree very well with the experimental data. The fluence-dependent $\tau_1$ further confirms the effectiveness of TTM (see Supplemental Material \cite{Suppl}). Employing $\tau^{-1}=3\hbar\lambda\langle\omega^{2}\rangle(\pi k_{B}T_{e})^{-1}$ \cite{Allen PRL 1987}, we can further obtain the dimensionless e-ph coupling constant $\lambda$ to be $\sim$0.17 via $\tau_{1}(T)$. During the calculations, we obtained $\lambda\langle\omega^{2}\rangle\simeq1.8$ $\times 10^{25}$~Hz$^{2}\simeq$78.2~meV$^2$, where the second moment of the phonon spectrum $\langle\omega^{2}\rangle$ is $\sim$6.2$\times 10^{4}$~K$^{2}$ extracted via the Debye model.
 
\begin{figure}
	\includegraphics[width=9cm]{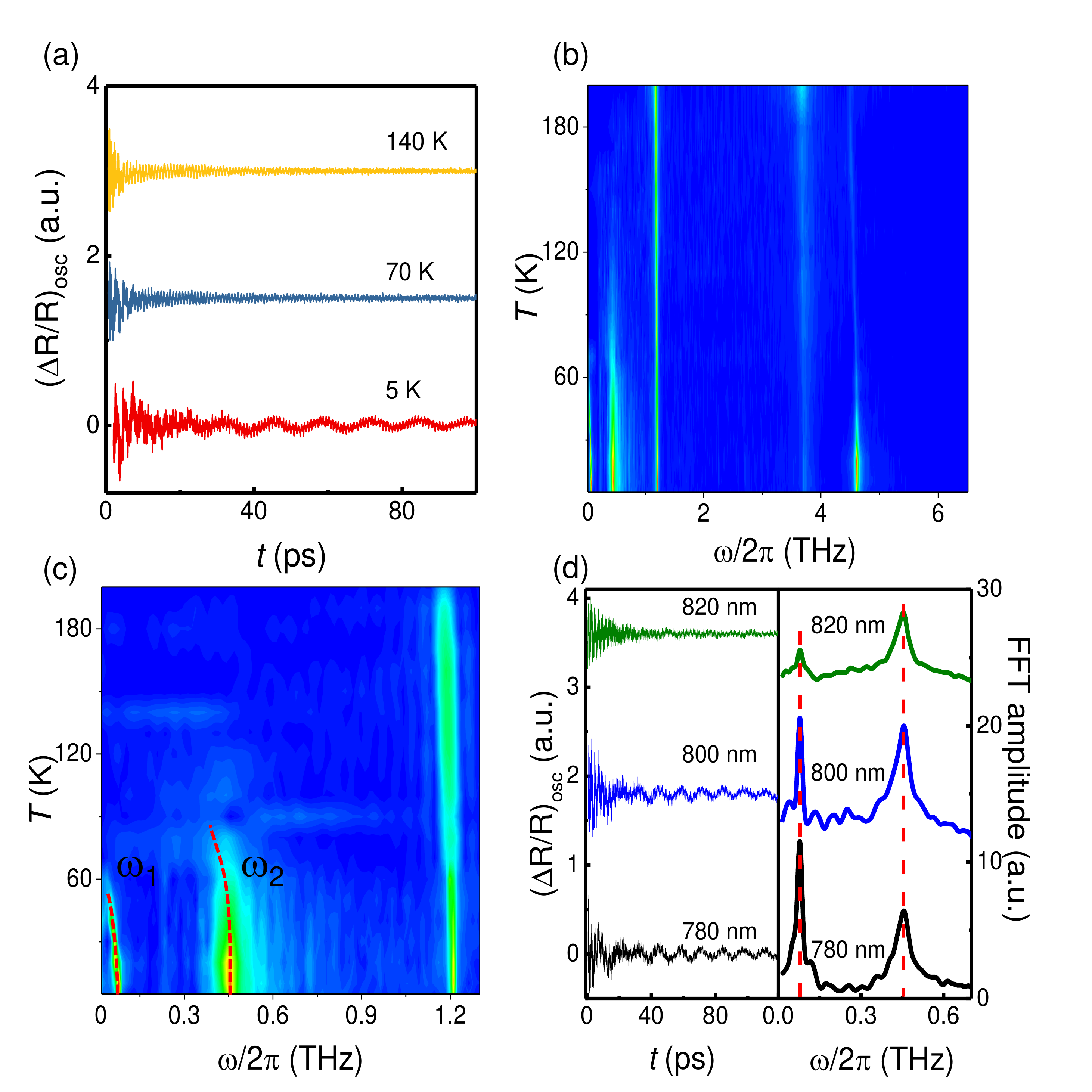}
	\caption{\label{fig:oscillation} (a) Extracted oscillations for three typical temperatures. (b) The temperature-dependent Fourier transform spectra for the extracted oscillations. (c) The two low energy modes, $\omega_1$ and $\omega_2$, as a function of temperature in the frequency domain. The red dashed curves are guide lines. (d) At 5 K, wavelength dependence of the two low energy modes, $\omega_1$ and $\omega_2$, indicated by the red dashed lines.}
	\vspace*{-0.2cm}
\end{figure}

The $\tau_{2}$ decay process, with a timescale of $\sim$3-18 ps, illustrates a totally different $T$-dependent behavior as that of the $\tau_1$ process (see Fig.~\ref{fig:decay}(c)). After the e-ph thermalization, the nonequilibrium electrons (holes) can still accumulate in the conduction (valence) bands. We might thus consider the phonon-assisted e-h recombination \cite{She PRB 2013,Dai PRB 2015}, where the electron and hole recombine with the assistance of e-ph scattering between the electron and hole pockets, as illustrated in Fig.~\ref{fig:decay}(d). In fact, the conduction and valence band extrema for ZrTe$ _{5} $ are in close proximity within the Brillouin zone \cite{Weng_PRX_2014,Monserrat B Physical Review Research 2019}, in favor of such interband e-h scatterings. Therefore, we assign the $\tau_{2}$ decay to this type of recombination process. This inference is not only consistent with the fluence ($F$) dependence of $\tau_2^{-1}$ (inset of Fig.~2(c)) \cite{D PRL 2010,Liu PRL 2020}, but also is strongly justified by the excellent agreement between the experimental $\tau_2$ and the theoretical fitted results based on such recombination \cite{Suppl}, as shown in the Fig.~\ref{fig:decay}(c).

The peculiar $\tau_{s}$ process emerges below $T^*$, implying that the $\tau_{s}$ decay should be associated with the physics behind the transition. At first glance, $T$-dependent behavior of $\tau_{s}$ well resembles that of $\tau_{1}(T)$. So we suspect that this $\tau_{s}$ process should microscopically resemble the e-ph scattering but involves unknown bosonic excitations emerging below $T^*$. Numerically, we still can use $\tau^{-1}=3\hbar\lambda\langle\omega^{2}\rangle(\pi k_{B}T_{e})^{-1}$ to fit $\tau_{s}$ based on the TTM \cite{Suppl}. However, the fitted parameter $\lambda\langle\omega^{2}\rangle$ ($\sim$5.47 meV$^2$) only gives an upper limit because the TTM now evolves into a three temperature model \cite{Wang PRB 2018}. At this stage, although the type of boson associated with $\tau_{s}$ is unknown (to be identified later), we can tell its energy is not very high and probably close to scale of the acoustic phonons based on its timescale (1-2 ps) \cite{Gedik PRL2011}.   

One may argue whether the $\tau_{s}$ process is associated with the Fermi level shifting from the valence band to the conduction band as the temperature decreases across $T^*$ \cite{Zhang NC 2017}. However, previous ultrafast optical studies on graphene \cite{I PRB 2020} demonstrated that when the Fermi level was continuously tuned across the Dirac point, no emergence or disappearance of any quasiparticle decay channel was observed except for change of the e-ph relaxation time. Therefore, this effect should be irrelevant here. 

\begin{figure}
	\includegraphics[width=9cm]{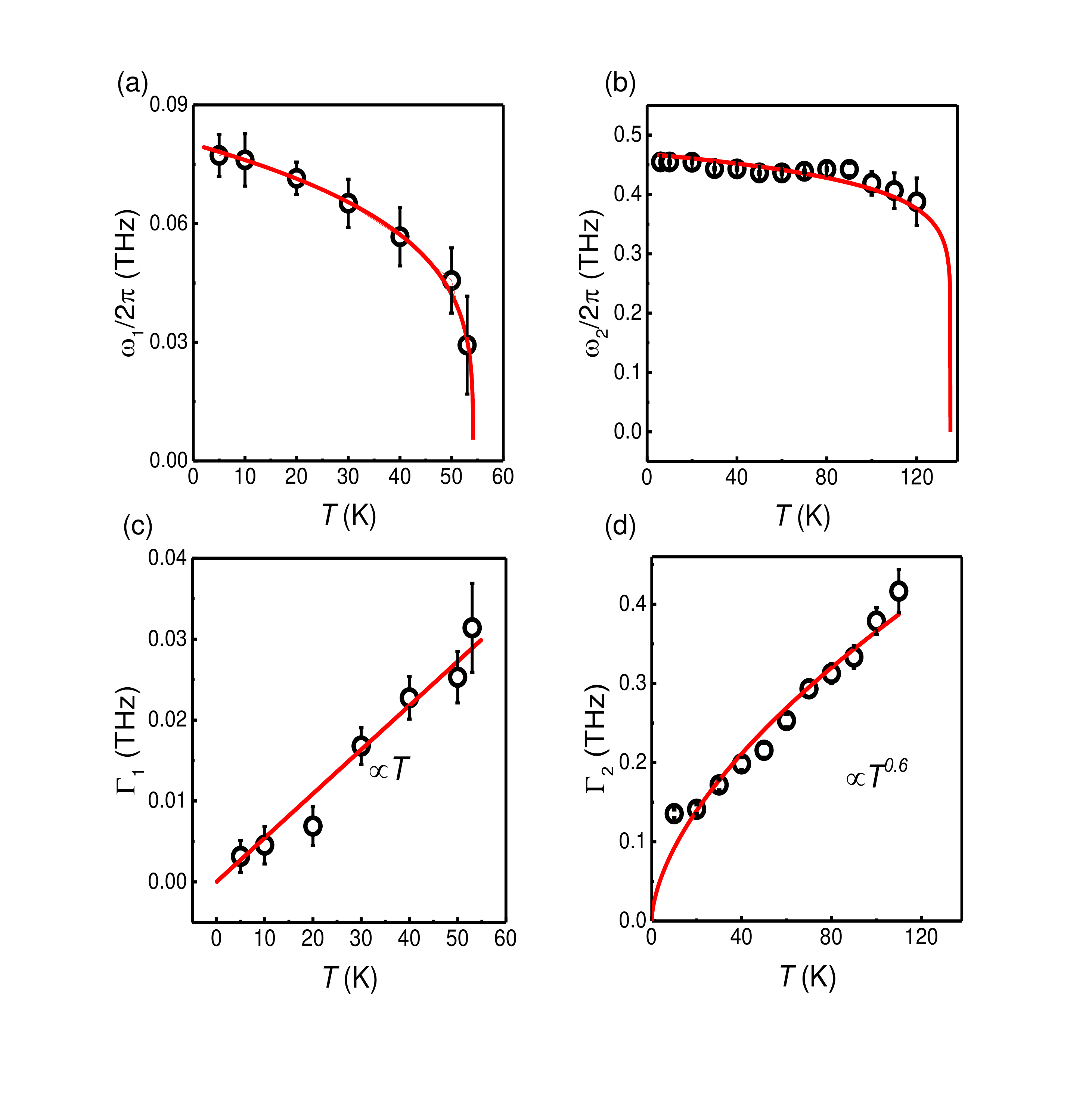}
	\caption{\label{fig:cdw}  $T$-dependence of the (a) $\omega_1$ and (b) $\omega_2$ modes. The red lines are fit described in the main text. (c) and (d) are $T$-dependent damping rates of the amplitudon for $\omega_1$ and $\omega_2$ modes, respectively. The red lines are fits using a power law ($\propto T^\alpha$).}
	\vspace*{-0.2cm}
\end{figure}

Now let us focus on the “rich" oscillations observed in $\Delta R(t)/R$. Fig.~\ref{fig:oscillation}(a) shows the typical time domain signals at temperatures below and above $T^*$. The temperature evolution of the oscillatory signals can be revealed by the Fourier transform data in the frequency domain (see Fig.~\ref{fig:oscillation}(b)), where five modes with distinct frequencies are clearly observed. 

In general, the oscillations with THz frequency in pump-probe spectroscopy arise from the coherent optical phonon modes near the $\Gamma$ point due to the coherent Raman scattering or displacive excitation \cite{H PRB 1992}. The exact modes can be revealed by a direct comparison with the Raman spectroscopy data \cite{Taguchi I Solid state communications 1983,Landa G Solid state communications 1984} (see also the Supplemental Material \cite{Suppl}). We thus obtain that three modes with center frequencies near $\sim$1.2~THz, $\sim$3.7~THz, and $\sim$4.6~THz are attributed to the $A_{g}$ phonon modes. These modes are characteristics of a high quality ZrTe$_5$ single crystal, first-time observed simultaneously in the time-resolved measurements. Temperature-dependence of all these modes can be well described by the anharmonic phonon model \cite{Suppl}.

Unexpectedly, as shown in Figs.~\ref{fig:oscillation}(b)-(d), we also clearly observed two low energy modes with frequencies of $\omega_{1}/2\pi\sim$0.08~THz and $\omega_{2}/2\pi\sim$0.45~THz, respectively. These two modes cannot be attributed to the optical phonons since their frequencies are well under the low frequency limit of the Raman-active optical phonons, i.e. $\sim$1~THz, obtained by our theoretical calculations \cite{Suppl}. In specific, the $\omega_{1}$ mode, emerging below $\sim$60 K, exhibits a clear softening as the temperature increases. The $\omega_{2}$ mode is non-observable above $\sim T^*$. One might first suspect they originate from the low energy acoustic phonons. However, the estimated phonon frequency and the corresponding probe-wavelength dependent shift significantly deviate from our observations (see Fig.~\ref{fig:oscillation}(d) and Supplemental Material \cite{Suppl}). Therefore, we can exclude such a possibility \cite{J PRB 2006,D PRB 2005, R PRL 2017}.   

Since in ZrTe$_5$ the long-range CDW order stabilized using the strong external magnetic field has been proposed to explain the quantum Hall effect \cite{Tang Nature 2019,Qin_PRL_2020}, these two modes, thus, might be attributed to the CDW instabilities triggered by the low-energy acoustic phonon mode \cite{G book 1994}. The dynamical CDW may give rise two collective excitations: (1) the amplitude mode (amplitudon) exhibits a gap when the wave vector approaches zero, and (2) the phase mode (phason) is gapless. Because the phason has similar dispersion relation as that of acoustic phonon, it is expected to bear the same wavelength dependence \cite{R PRL 2017} and thus cannot contribute to these two low energy modes. We thus finally arrive at the most reasonable candidate for these two modes, i.e. the amplitudon of CDW. 

Particularly, we notice that $T$ dependence of these two modes is quite similar with the temperature evolution of the fluctuating CDW \cite{G book 1994,R PRL 2017,Yusupov_PRL_2008}, as seen in Figs.~\ref{fig:cdw}(a) and (b). $\omega_{1}$ and $\omega_{2}$ as a function $T$ can well be fitted by a mean-field-like $T$ dependence \cite{Yusupov_PRL_2008,G book 1994}: $\omega_{A}=\omega_{A}(0)(1-T/T_C)^\beta$, where $T_C$ is defined as a transition temperature. Satisfactory fits to the data in Figs.~\ref{fig:cdw}(a)-(b) give \cite{note2}: $\beta_{\omega_1}\simeq$0.25 and $\beta_{\omega_2}\simeq$0.1 corresponding to $T_C^{\omega_1}\simeq54$~K and $T_C^{\omega_2}\simeq135$~K, respectively. $\beta_{\omega_1}$ having a value of $\sim$1/4 indicates that $\omega_1$ mode has a pure one-dimensional origin \cite{note2}. In fact, such character is consistent with our theoretical calculations \cite{Suppl}. Specifically, the obtained phonon linewidth in ZrTe$_5$ has a quasi-one-dimensional character, i.e. the corresponding values of three acoustic modes along the $\Gamma-Z$ direction (inter-plane) are in general larger than those along others. Since the e-ph interactions dominate the order parameter \cite{Qin_PRL_2020}, it is not surprising that a fluctuating CDW can emerge along the inter-plane direction ($b$-axis). By contrast, the $\omega_2$ mode cannot belong to a simple one-dimensional case. The intricate $\omega_2$ mode might have higher dimensional characteristics, and needs to be unveiled by further investigations. 

The $T$-dependent damping rate $\Gamma_1$ ($\Gamma_2$) for the $\omega_1$ ($\omega_2$) mode can be extracted by fitting the oscillatory signals via a damped sine function \cite{Suppl}, as shown in Figs.~\ref{fig:cdw}(c) and (d). In order to compare with the theoretical calculations \cite{Takada_PRB_1985,Richard_SSC_1993}, we used a power law ($\propto T^\alpha$) to fit the data, and obtained that $\Gamma_1$ and $\Gamma_2$ scale well as $T$ and $T^{0.6}$, respectively. Such $T$-scalings clearly deviate from the theoretical intrinsic damping \cite{Takada_PRB_1985,Richard_SSC_1993,Torchinsky_NM_2013}, i.e. $T^2$ away from $T_{C}$ and $T^5$ close to $T_{C}$, respectively. Indeed, decay of the amplitudon includes both the intrinsic damping and correlation time of the fluctuating CDW. Therefore, $\Gamma_{1}^{-1}$ and $\Gamma_{2}^{-1}$ here provide a lower bound of the correlation time $(\tau_F)_{1}$ and $(\tau_F)_{2}$. 

As is known, the acoustic phonon may soften significantly at $q=2k_{f}$ upon the phase transition of CDW \cite{G book 1994}. In the $q=$0 limit, $\omega_{A}\simeq\lambda^{1/2}\omega_{2k_{f}}$, where $\omega_{2k_{f}}$ is the frequency of the softening phonon above the phase transition connected to the frequency of amplitudon ($\omega_{A}$) via the e-ph coupling constant $\lambda$. Therefore, using a linear dispersion approximation with sound velocity of $v_s\sim10^3$~m/s \cite{Wang_JAP_2018}, $k_f$ can be estimated to have values of 0.053~\AA$^{-1}$ and 0.34~\AA$^{-1}$ for $\omega_1$ and $\omega_2$ modes, respectively. Although we do not have accurate knowledge of anisotropic $v_s$ associated with $\omega_1$ and $\omega_2$ modes, based on these two estimated values we can conclude that their corresponding modulation wave vectors should be very small. Such uncommonly small $k_f$ values indicate that our observed density waves have tremendously long lattice modulation periods in the real space. These results are directly due to the extremely low energies of $\omega_1$ and $\omega_2$ modes. Interestingly, these derived values are consistent with the work demonstrating the 3D quantum Hall effect \cite{Tang Nature 2019}.   

Since $\omega_1$ and $\omega_2$ modes now can be attributed to the fluctuating CDW that obeys the Bose-Einstein statistics, they can naturally be assigned to the bosons contributing to the $\tau_s$ decay process emerging below $T^*$. However, the $\omega_2$ mode is the mostly possible candidate due to two reasons: (1) appearance temperature of this mode coincides with that of the $\tau_s$ process; (2) No clear anomalies at $T_C^{\omega_1}$ were found in the $T$-dependent $\tau_s$ and $A_s$ \cite{Suppl}; (3) energy of the $\omega_1$ mode is nearly smaller by an order than that of the $\omega_2$ mode so that the timescale of energy exchange between the nonequilibrium quasiparticles and $\omega_1$ mode can mix up with the long thermal diffusion process, where large number of low-energy acoustic phonons will overwhelm the CDW modes.    

In summary, using the ultrafast optical spectroscopy we present a detailed investigation on the quasi-one-dimensional system ZrTe$_5$. Our results reveal that below $T^*$, there emerges a new quasiparticle decay process with a timescale of $\sim$1-2 ps, in addition to the two relaxation dynamics persisting at all temperatures with timescales of $\sim$0.15-0.3 ps and $\sim$3-18 ps, respectively. We address that this new decay process is most likely due to appearance of some novel collective excitations in ZrTe$_5$, while the other two can be attributed to the electron-phonon (e-ph) scattering and the phonon-assisted electron-hole (e-h) recombination processes. Surprisingly, our experimental data for the first time unambiguously reveal two coherent oscillations with extremely small energies occurring below two distinct temperatures $\sim$54 K and $\sim$135 K, respectively. We argue that they arise from the amplitude mode of fluctuating CDW triggered by the acoustic phonons, and can well explain the new observed quasiparticle decay below $T^*$. Whether these findings are connected to the other emergent phenomena in this system deserves further explorations.

We acknowledge the valuable discussion from Hrvoje Petek. This work was supported by the National Natural Science Foundation of China (Grants Nos. 11974070, 11734006, 11925408, 11921004), the Frontier Science Project of Dongguan (2019622101004), the National Key R\&D Program of China (Grant Nos. 2016YFA0300600 and 2018YFA0305700), the Strategic Priority Research Program of Chinese Academy of Sciences (Grant No. XDB33000000), the K. C. Wong Education Foundation (Grant No. GJTD-2018-01), the Beijing Natural Science Foundation (Grant No. Z180008), the Beijing Municipal Science and Technology Commission (Grant No. Z191100007219013), and the CAS Interdisciplinary Innovation Team.


\begin{thebibliography}{text}
\bibitem{G book 1994} G. Grüner, \textit{Density waves in solids} (Perseus, Cambridge, MA, 1994).
\bibitem{Kivelson_RMP_2003} S. A. Kivelson, I. P. Bindloss, E. Fradkin, V. Oganesyan, J. M. Tranquada, et al., Rev. Mod. Phys. \textbf{75}, 1201 (2003).
\bibitem{Weng_PRX_2014} H. Weng, X. Dai, and Z. Fang, Phys. Rev. X \textbf{4}, 011002 (2014).
\bibitem{J sciadv 2019}	J. Mutch, W. C. Chen, P. Went, T. Qian, I. Z. Wilson, et al., Sci. Adv. \textbf{5}, eaav9771 (2019).
\bibitem{C PRX 2020} C. Vaswani, L. L. Wang, D. H. Mudiyanselage, Q. Li, P. M. Lozano, et al., Phys. Rev. X \textbf{10}, 021013 (2020).
\bibitem{R PRX 2016}R. Wu, J. Z. Ma, S. M. Nie, L. X. Zhao, X. Huang, et al., Phys. Rev. X \textbf{6}, 021017 (2016).
\bibitem{G PRL 2016}G. Manzoni, L. Gragnaniello, G. Autès, T. Kuhn, A. Sterzi, et al., Phys. Rev. Lett. \textbf{117}, 237601 (2016).
\bibitem{H PRB 2017}H. Xiong, J. A. Sobota, S. L. Yang, H. Soifer, A. Gauthier, et al., Phys. Rev. B \textbf{95}, 195119 (2017).
\bibitem{Y PRB 2018}Y. Y. Lv, B. Bin Zhang, X. Li, K. W. Zhang, X. B. Li, et al., Phys. Rev. B \textbf{97}, 115137 (2018).
\bibitem{RY PRB 2015}R. Y. Chen, S. J. Zhang, J. A. Schneeloch, C. Zhang, Q. Li, et al., Phys. Rev. B \textbf{92}, 075107 (2015).
\bibitem{Y NC 2016}Y. Liu, X. Yuan, C. Zhang, Z. Jin, A. Narayan, et al., Nat. Commun. \textbf{7}, 12516 (2016).
\bibitem{Wang 2018}J. Wang, J. Niu, B. Yan, X. Li, R. Bi, et al., Proc. Natl. Acad. Sci. U. S. A. \textbf{115}, 9145 (2018).
\bibitem{Terry PRB 1999}T. M. Tritt, N. D. Lowhorn, R. T. Littleton, A. Pope, C. R. Feger, and J. W. Kolis, Phys. Rev. B \textbf{60}, 7816 (1999).
\bibitem{Niu PRB2017}J. Niu, J. Wang, Z. He, C. Zhang, X. Li, et al., Phys. Rev. B \textbf{95}, 035420 (2017).
\bibitem{Jiang PRB 2017}Y. Jiang, Z. L. Dun, H. D. Zhou, Z. Lu, K. W. Chen, et al., Phys. Rev. B \textbf{96}, 041101 (2017).
\bibitem{ZG PNAS 2017}Z. G. Chen, R. Y. Chen, R. D. Zhong, J. Schneeloch, C. Zhang, et al., Proc. Natl. Acad. Sci. U. S. A. \textbf{114}, 816 (2017).
\bibitem{B PRL 2018}B. Xu, L. X. Zhao, P. Marsik, E. Sheveleva, F. Lyzwa, et al., Phys. Rev. Lett. \textbf{121},  187401 (2018).
\bibitem{Jiang PRL2020} Y. Jiang, J. Wang, T. Zhao, Z. L. Dun, Q. Huang, et al., Phys. Rev. Lett. \textbf{125}, 046403 (2020).
\bibitem{Li PRL 2016}X. B. Li, W. K. Huang, Y. Y. Lv, K. W. Zhang, C. L. Yang, et al., Phys. Rev. Lett. \textbf{116}, 176803 (2016).
\bibitem{Li_NPhy_2016} Q. Li, D. E. Kharzeev, C. Zhang, Y. Huang, I. Pletikosić, et al., Nat. Phys. \textbf{12}, 550 (2016).
\bibitem{Liang NP 2018} T. Liang, J. Lin, Q. Gibson, S. Kushwaha, M. Liu, et al., Nat. Phys. \textbf{14}, 451 (2018).
\bibitem{Tang Nature 2019}F. Tang, Y. Ren, P. Wang, R. Zhong, J. Schneeloch, et al., Nature \textbf{569}, 537 (2019).
\bibitem{FJ PRB 1981}F. J. DiSalvo, R. M. Fleming, and J. V. Waszczak, Phys. Rev. B \textbf{24}, 2935 (1981).
\bibitem{Qin_PRL_2020} F. Qin, S. Li, Z. Z. Du, C. M. Wang, W. Zhang, et al., Phys. Rev. Lett. \textbf{125}, 206601 (2020).
\bibitem{Fu_PRL_2020}B. Fu, H. W. Wang, and S. Q. Shen, Phys. Rev. Lett. \textbf{125}, 256601 (2020).


\bibitem{Gedik PRL2011}D. Hsieh, F. Mahmood, J. W. McIver, D. R. Gardner, Y. S. Lee, and N. Gedik, Phys. Rev. Lett. \textbf{107}, 077401 (2011).
\bibitem{Wang PRL 2016}	M. C. Wang, S. Qiao, Z. Jiang, S. N. Luo, and J. Qi, Phys. Rev. Lett. \textbf{116}, 036601 (2016).

\bibitem{Yusupov_PRL_2008}R. V. Yusupov, T. Mertelj, J. H. Chu, I. R. Fisher, and D. Mihailovic, Phys. Rev. Lett. \textbf{101}, 246402 (2008).
\bibitem{Torchinsky_NM_2013} D. H. Torchinsky, F. Mahmood, A. T. Bollinger, I. Božović, and N. Gedik, Nat. Mater. \textbf{12}, 387 (2013).
\bibitem{Liu PRL 2020}Y. P. Liu, Y. J. Zhang, J. J. Dong, H. Lee, Z. X. Wei, et al., Phys. Rev. Lett. \textbf{124}, 057404 (2020).

\bibitem{note1} The unique properties of ZrTe$_5$ do not manifest them in previous time-resolved optical experiments \cite{Zhang PRB 2019,Li PRB 2020}, primarily due to the critical deficiencies in these measurements such as the usage of unreliable samples, evidenced by the absence of various coherent Raman modes characterizing the quality of ZrTe$_5$. 





\bibitem{Zhang PRB 2019}X. Zhang, H. Y. Song, X. C. Nie, S. B. Liu, Y. Wang, et al., Phys. Rev. B \textbf{99}, 125141 (2019).
\bibitem{Li PRB 2020}N. Li, W. Liang, and S. N. Luo, Phys. Rev. B \textbf{101}, 014304 (2020).
\bibitem{Suppl} See Supplemental Material for additional data, more experimental and theoretical details, which includes Refs.~\cite{Weng_PRX_2014,Jiang PRB 2017,Li_NPhy_2016,1,4,5,6,7,9,Landa G Solid state communications 1984,10,11,12,Qi PRL 2013,Balkanski M PRB 1983,Menendez J PRB 1984,Tang H PRB 1991,17,18,19,20,22,23,24,25,26,27,28}.
\bibitem{Qi PRL 2013} J. Qi, T. Durakiewicz, S. A. Trugman, J. X. Zhu, P. S. Riseborough, et al., Phys. Rev. Lett. \textbf{111}, 057402 (2013).
\bibitem{1} T. Dekorsy, G. C. Cho, and H. Kurz, \textit{in Light Scattering in Solids VIII,} Ed. by M. Cardona and G. Güntherodt, (Springer,
Berlin, 2000)
\bibitem{4}P. B. Allen, Phys. Rev. B \textbf{6}, 2577 (1972).
\bibitem{5}R. H. Groeneveld, R. Sprik, and A. Lagendijk, Phys. Rev. B \textbf{51}, 11433 (1995).
\bibitem{6}A. A. Lopez, Phys. Rev. \textbf{175}, 823 (1968).
\bibitem{7}A. Zwick, G. Landa, R. Carles, M. A. Renucci, and A. Kjekshus, Solid State Commun \textbf{44}, 89 (1982).
\bibitem{9}R. Merlin, Solid State Commun. \textbf{102}, 207 (1997).
\bibitem{10}G. A. Garrett, T. F. Albrecht, J. F. Whitaker, and R. Merlin, Phys. Rev. Lett. \textbf{77}, 3661 (1996).
\bibitem{11}I. Gdor, T. Ghosh, O. Lioubashevski, and S. Ruhman, J. Phys. Chem. Lett. \textbf{8}, 1920 (2017).
\bibitem{12}G. Batignani, C. Ferrante, G. Fumero, and T. Scopigno, J. Phys. Chem. Lett. \textbf{10}, 7789 (2019).
\bibitem{17} Z. Guo, H. Gu, M. Fang, B. Song, W. Wang, X. Chen, C. Zhang, H. Jiang, L. Wang, and S. Liu, ACS Materials Lett. \textbf{3}, 525 (2021).
\bibitem{18}V. L. Ginzburg, Sov. Phys. Usp. \textbf{5}, 649 (1963).
\bibitem{19}Y. Toda, F. Kawanokami, T. Kurosawa, M. Oda, I. Madan, T. Mertelj, V. V. Kabanov, and D. Mihailovic, Phys. Rev. B \textbf{90}, 094513(2014).
\bibitem{20}A. Kogar, A. Zong, P. E. Dolgirev, X. Z. Shen, et al., Nat. Phys. \textbf{16}, 159 (2019).
\bibitem{22}S. Furuseth, L. Brattas, and A. Kjekshus. J. Acta Chemica Scandinavica \textbf{27}, 2367 (1973).
\bibitem{23}P. Giannozzi, S. Baroni, et al., J. Phys. Condens. Matter \textbf{21}, 395502 (2009).
\bibitem{24} P. Giannozzi, O. Andreussi, et al., J. Phys. Condens. Matter \textbf{29}, 465901 (2017).
\bibitem{25}S. Baroni, S. d. Gironcoli, A. D. Corso, and P. Giannozzi, Rev. Mod. Phys. \textbf{73}, 515 (2001).
\bibitem{26}D. R. Hamann. Phys. Rev. B \textbf{88}, 085117 (2013).
\bibitem{27}F. Giustino, M. L. Cohen, and S. G. Louie. Phys. Rev. B \textbf{76}, 165108 (2007).
\bibitem{28}S. Poncé, E.R. Margine, C. Verdi, and F. Giustino. Comput. Phys. Commun. \textbf{209}, 116 (2016).
\bibitem{Landa G Solid state communications 1984}G. Landa, A. Zwick, R. Carles, M. A. Renucci, and A. Kjekshus, Solid State Commun. \textbf{49}, 1095 (1984).

\bibitem{Balkanski M PRB 1983}M. Balkanski, R. F. Wallis, and E. Haro, Phys. Rev. B \textbf{28}, 1928 (1983).
\bibitem{Menendez J PRB 1984}J. Menéndez and M. Cardona, Phys. Rev. B \textbf{29}, 2051 (1984).

\bibitem{Tang H PRB 1991} 	H. Tang and I. P. Herman, Phys. Rev. B \textbf{43}, 2299 (1991).

\bibitem{W J 2016}W. Yu, Y. Jiang, J. Yang, Z. L. Dun, H. D. Zhou, et al., Sci. Rep. \textbf{6}, 353357 (2016).
\bibitem{Qi APL 2010}J. Qi, X. Chen, W. Yu, P. Cadden-Zimansky, D. Smirnov, et al., Appl. Phys. Lett. \textbf{97}, 182102 (2010).
\bibitem{O PRL 2013}J. P. Hinton, J. D. Koralek, G. Yu, E. M. Motoyama, Y. M. Lu, et al., Phys. Rev. Lett. \textbf{110}, 217002 (2013).



\bibitem{D PRL 2002}D. J. Hilton and C. L. Tang, Phys. Rev. Lett. \textbf{89}, 146601 (2002).






\bibitem{J Book 2013}	J. Shah, \textit{Ultrafast spectroscopy of semiconductors and semiconductor nanostructures}. (Springer Science  Business Media, 2013).
\bibitem{D Rev Mod Phys}D. N. Basov, R. D. Averitt, D. Van Der Marel, M. Dressel, and K. Haule, Rev. Mod. Phys. \textbf{83}, 471 (2011).
\bibitem{R 2002}R. D. Averitt and a J. Taylor, J. Phys. Condens. Matter \textbf{14}, R1357 (2002).
\bibitem{R PRB 1995}R. H. M. Groeneveld, R. Sprik, and A. Lagendijk, Phys. Rev. B \textbf{51}, 11433 (1995).
\bibitem{M PRB 2005}M. Hase, K. Ishioka, J. Demsar, K. Ushida, and M. Kitajima, Phys. Rev. B \textbf{71}, 184301 (2005).
\bibitem{M Nano Lett 2012} M. Hajlaoui, E. Papalazarou, J. Mauchain, G. Lantz, N. Moisan, et al., Nano Lett. \textbf{12}, 3532 (2012).


\bibitem{Liang APL 2014}L. Cheng, C. La-O-Vorakiat, C. S. Tang, S. K. Nair, B. Xia, et al., Appl. Phys. Lett. \textbf{104}, 211906 (2014).



\bibitem{Allen PRL 1987}P. B. Allen, Phys. Rev. Lett. \textbf{59}, 1460 (1987).
\bibitem{She PRB 2013}Y. M. Sheu, Y. J. Chien, C. Uher, S. Fahy, and D. A. Reis, Phys. Rev. B \textbf{87}, 075429 (2013).
\bibitem{Dai PRB 2015} Y. M. Dai, J. Bowlan, H. Li, H. Miao, S. F. Wu, et al., Phys. Rev. B \textbf{92}, 161104 (2015).

\bibitem{Monserrat B Physical Review Research 2019}B. Monserrat and A. Narayan, Phys. Rev. Res. \textbf{1}, 033181 (2019).



\bibitem{D PRL 2010}D. H. Torchinsky, G. F. Chen, J. L. Luo, N. L. Wang, and N. Gedik, Phys. Rev. Lett. \textbf{105}, 027005 (2010).
\bibitem{Wang PRB 2018} M. C. Wang, H. S. Yu, J. Xiong, Y. F. Yang, S. N. Luo, et al., Phys. Rev. B \textbf{97}, 155157 (2018).
\bibitem{Zhang NC 2017}Y. Zhang, C. Wang, L. Yu, G. Liu, A. Liang, et al., Nat. Commun. \textbf{8}, 15512 (2017).

\bibitem{I PRB 2020}I. Katayama, K. I. Inoue, Y. Arashida, Y. Wu, H. Yang, et al., Phys. Rev. B \textbf{101}, 245408 (2020).




\bibitem{H PRB 1992} H. J. Zeiger, J. Vidal, T. K. Cheng, E. P. Ippen, G. Dresselhaus, and M. S. Dresselhaus, Phys. Rev. B \textbf{45}, 768 (1992).

\bibitem{Taguchi I Solid state communications 1983}I. Taguchi, A. Grisel, and F. Levy, Solid State Commun. \textbf{46}, 299 (1983).

\bibitem{D PRB 2005}D. Lim, V. K. Thorsmølle, R. D. Averitt, Q. X. Jia, K. H. Ahn, et al., Phys. Rev. B \textbf{71}, 134403 (2005).
\bibitem{J PRB 2006}J. K. Miller, J. Qi, Y. Xu, Y. J. Cho, X. Liu, et al., Phys. Rev. B \textbf{74}, 113313 (2006).

\bibitem{R PRL 2017}R. Y. Chen, S. J. Zhang, M. Y. Zhang, T. Dong, and N. L. Wang, Phys. Rev. Lett. \textbf{118}, 107402 (2017).

\bibitem{note2} $\beta$ will arrive at 1/4 under the condition of $\omega_A\gg$$v_fq$ in one-dimensional case \cite{G book 1994}.

\bibitem{Takada_PRB_1985} S. Takada, K. Y. M. Wong, and T. Holstein, Phys. Rev. B \textbf{32}, 4639 (1985).

\bibitem{Richard_SSC_1993} J. Richard and J. Chen, Solid State Commun. 86, 485 (1993).

\bibitem{Wang_JAP_2018}C. Wang, H. Wang, Y. B. Chen, S. H. Yao, and J. Zhou, J. Appl. Phys. \textbf{123}, 175104 (2018).





\end{thebibliography}
\end{document}